\documentclass[english]{iopart}

\usepackage{iopams}
\usepackage{babel}
\usepackage{units}
\usepackage{amsbsy}
\usepackage{amssymb}
\usepackage{graphicx}
\usepackage{esint}

\usepackage{hyperref}

\makeatother

\begin{document}

\title[]{Interaction of a Bose-Einstein Condensate and a Superconductor via Eddy Currents}

\author{Igor Sapina$^{1}$ and Thomas Dahm$^{1,2,3}$}

\address{$^{1}$Institut f\"ur Theoretische Physik and Center for Collective
Quantum Phenomena, Universit\"at T\"ubingen, Auf der Morgenstelle 14,
D-72076 T\"ubingen, Germany}

\address{$^{2}$Fakult\"at f\"ur Physik, Universit\"at Bielefeld, Postfach 100131,
D-33501 Bielefeld, Germany}

\address{$^{3}$Department of Physics, University of Hull, Hull, HU6 7RX,United
Kingdom}

\ead{isapina@physik.uni-bielefeld.de}

\ead{thomas.dahm@uni-bielefeld.de}

\begin{abstract}
We study center-of-mass oscillations of a dipolar Bose-Einstein condensate
in the vicinity of a superconducting surface. We show that the magnetic
field of the magnetic dipoles induces eddy currents in the superconductor,
which act back on the Bose-Einstein condensate. This leads to a shift
of its oscillation frequency and to an anharmonic coupling of the
Bose-Einstein condensate with the superconductor. The anharmonicity
creates a coupling to one of the collective modes of the condensate
that can be resonantly enhanced, if the parameters of the condensate
are chosen properly. This provides a new physical mechanism to couple
a Bose-Einstein condensate and a superconductor which becomes significant
for $^{52}$Cr, $^{168}$Er or $^{164}$Dy condensates in superconducting
mircotraps.
\end{abstract}


\maketitle

\section{Introduction}

In the last years hybrid quantum systems have come into focus of research
in the context of quantum information processing \cite{Kubo,Amsuess,OConnell,Wallraff}.
They are able to combine the strengths of qubits based on solid state
devices, which can be better controlled, and qubits based on atomic
systems, which promise longer coherence times. Such a hybrid can
consist of a superconductor coupled to a Bose-Einstein condensate
(BEC). To achieve a controlled coupling between a superconductor and
a BEC it is necessary to understand the interaction between those
two systems. 

Coupled quantum systems based on BECs and solid state devices have
been suggested theoretically \cite{Fleischhauer,Verdu,Folman}. While
the influence of a solid on a BEC is sizable and has been studied
well in the past, normally the influence of a BEC on a solid is very
weak due to the low density of the BEC. In recent years, BECs have
been condensed in superconducting microtraps \cite{Nirrengarten,Mukai,CanoPRL,Siercke,Zhang}
allowing a close approach of a BEC to a superconducting surface \cite{Kasch,Markowsky,Dikovsky,Sokolovsky}.
The use of a superconducting microtrap as opposed to a metallic one
allows for a significantly longer lifetime of the atomic cloud in
close vicinity of the surface \cite{Kasch,Skagerstam,Hohenester,Fermani},
and therefore also longer coherence times. It also promises a successful
coupling of these two macroscopic quantum phenomena \cite{Verdu,Folman,Bensky,Salem,CanoEPJD,Bernon}.

In the last decade there has also been intensive research, theoretical
as well as experimental, in the field of dipolar BECs. The high magnetic
dipole moment of the atoms leads to a long ranged anisotropic interaction
between the atoms in a BEC. This interaction is responsible for a
number of interesting phenomena observable in dipolar BECs \cite{Lahaye}.
For some time $^{52}$Cr \cite{Chrom1} has been the only experimentally
realized dipolar BEC. But most recently also the condensation of $^{168}$Er
and $^{164}$Dy \cite{Erbium,Dysprosium} has been achieved.

Here we will study a dipolar BEC in close vicinity of a superconductor.
We suggest that the mutual interaction of a dipolar BEC with a superconductor
can become sizeable due to the large magnetic dipole moment of the
atoms. Specifically, center-of-mass oscillations of the dipolar condensate
within its trap create eddy currents in the superconductor surface.
These eddy currents, in turn, shift the oscillation frequency of the
condensate. Their anharmonicity creates a coupling of the center-of-mass
oscillations of the condensate with one of its collective modes. We
show that this anharmonic coupling can be resonantly enhanced, allowing
for a sizeable interaction of the BEC with the solid and a new mechanism
for coupling a BEC with a superconductor.

\section{Interaction between the superconducting surface and the Bose-Einstein
condensate}

\begin{figure}
\flushright\includegraphics{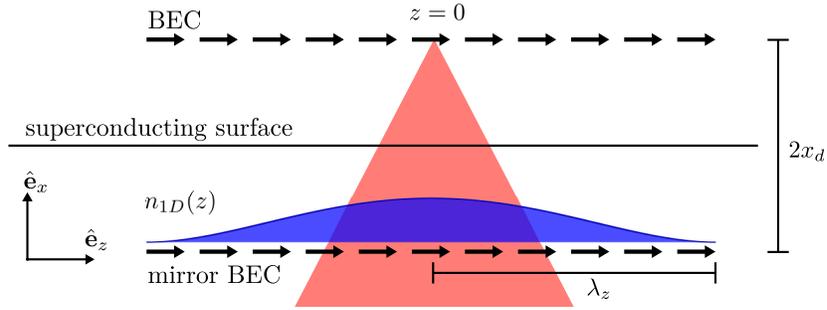}
\caption{\label{Fig01} The dipole at $z=0$ interacts with the mirror dipoles
distributed with $n_{1\mathrm{D}}\left(z\right)$ along the $z$-axis.
The interaction sign with the dipoles in the red region is negative
and with the dipoles beyond the red region its positive. The BEC has
an optimal length when all dipoles are in the red region.}
\end{figure}

Consider a weakly interacting dipolar BEC of $N$ atoms at temperature
$T=0$ confined by an external harmonic trapping potential $V_{T}\left(\mathbf{r}\right)$.
The trapping potential could be generated by the magnetic field of
a superconducting atom chip or a laser field, for example. The spins
are all aligned along $\hat{\mathbf{e}}_{z}$ by an external magnetic
field (see Fig. \ref{Fig01}). The interaction $U\left(\mathbf{r},\mathbf{r}^{\prime}\right)$
between two atoms consists of two contributions. One contribution
is the isotropic contact interaction $U_{s}\left(\mathbf{r},\mathbf{r}^{\prime}\right)=g_{s}\delta^{(3)}\left(\mathbf{r}-\mathbf{r}^{\prime}\right)$,
where $g_{s}=\frac{4\pi\hbar^{2}a_{s}}{M}$ gives the strength of
the interaction and is determined by the mass $M$ and the $s$-wave
scattering length $a_{s}$. The other contribution is the long ranged
magnetic dipole-dipole interaction 
\begin{equation}
U_{\mathrm{md}}\left(\mathbf{r},\mathbf{r}^{\prime}\right)=-\,\frac{g_{D}}{4\pi}\left(\frac{3(z-z^{\prime})^{2}}{\left|\mathbf{r}-\mathbf{r}^{\prime}\right|^{5}}-\frac{1}{\left|\mathbf{r}-\mathbf{r}^{\prime}\right|^{3}}\right).\label{eq:Umd}
\end{equation}
 The strength is given by $g_{D}=\mu_{0}m^{2}$, where $m$ is the
magnetic dipole moment. The time evolution of the BEC is given by
the time dependent Gross-Pitaevskii equation (GPE) \cite{Pitaevskii&Stingari,Pethick&Smith}
\begin{equation}
\fl i\hbar\frac{\partial}{\partial t}\psi\left(\mathbf{r};\, t\right)=\left[-\frac{\hbar^{2}}{2M}\boldsymbol{\nabla}^{2}+V_{T}(\mathbf{r})+\left(N-1\right)\intop_{\mathbb{R}^{3}}\mathrm{d}\mathbf{r}^{\prime}\, U\left(\mathbf{r},\mathbf{r}^{\prime}\right)\left|\psi\left(\mathbf{r}^{\prime};\, t\right)\right|^{2}\right]\psi\left(\mathbf{r};\, t\right).\label{eq:TimeDepGPE}
\end{equation}
For simplicity we will model the superconductor by a superconducting
half space, which is a valid approximation when the BEC is sufficiently 
close to a plane superconductor surface. 
The presence of a superconducting half space modifies the magnetic
field distribution of a nearby magnetic dipole due to the currents
induced in the superconductor. As long as the distance between the
magnetic dipole and the surface of the superconductor (in our case
$\sim\unit[10]{\mu m}$) is larger than the magnetic penetration depth
of the superconductor ($\sim\unit[100]{nm}$ for Nb) and as long as
the oscillation frequency of the dipole motion (in our case $\sim\unit[1]{Hz}$-$\unit[1]{kHz}$)
is smaller than the gap frequency ($\sim\unit[100]{GHz}$ for Nb),
the superconductor acts as a perfect magnetic mirror. This means that
at the surface of a superconductor the normal component of a magnetic
induction field has to vanish $\mathbf{B}\cdot\hat{\mathbf{n}}=0$.
The field distribution of the magnetic dipole close to a superconductor
can thus be found by introducing a mirror dipole in the superconductor
and adding up the field of the dipole and the mirror dipole. The mirror
dipole emulates the effect of the induced eddy currents. This way
the magnetic interaction between a dipolar BEC and a superconductor
can be described by an additional external potential felt by the atoms
in the BEC due to the mirror BEC: 
\begin{equation}
U_{\mathrm{SC}}(\mathbf{r})=\intop_{\mathbb{R}^{3}}\mathrm{d}\mathbf{r}^{\prime}\, n(\mathbf{r}^{\prime})U_{\mathrm{md}}\left(\mathbf{r},\mathbf{r}^{\prime}\right).\label{eq:Usc}
\end{equation}
 Here, $n\left(\mathbf{r}^{\prime}\right)$ is the density distribution
of the mirror BEC. Note, that this potential depends on the number
of atoms in the BEC in contrast to other single-particle potentials
like the Casimir-Polder force for example.

The BEC ground state is the stationary solution of (\ref{eq:TimeDepGPE}),
which will be determined numerically below. Before we do that, let
us discuss first a useful approximation for the potential of the mirror.
With a sufficiently large number of atoms in the BEC the kinetic energy
can be neglected, which leads to the Thomas-Fermi (TF) approximation
\cite{Pitaevskii&Stingari,Pethick&Smith}. Within this approximation
an analytical expression for the density distribution $N\cdot\left|\psi\left(\mathbf{r}\right)\right|^{2}$
of a BEC can be given. In an harmonic potential the density distribution
has an ellipsoidal shape 
\begin{eqnarray}
n_{\mathrm{TF}}\left(\mathbf{r}\right) & =n_{0}\left(1-\frac{x^{2}}{\lambda_{x}^{2}}-\frac{y^{2}}{\lambda_{y}^{2}}-\frac{z^{2}}{\lambda_{z}^{2}}\right)\quad\mathrm{for}\,\mathbf{r}\in\mathbb{D}_{\mathrm{TF}}\label{eq:nTF}\\
\mathrm{with}\quad & \mathbb{D}_{\mathrm{TF}}=\left\{ \mathbf{r}\in\mathbb{R}^{3}\left|\frac{x^{2}}{\lambda_{x}^{2}}+\frac{y^{2}}{\lambda_{y}^{2}}+\frac{z^{2}}{\lambda_{z}^{2}}\leq1\right.\right\} \nonumber 
\end{eqnarray}
 and $n_{\mathrm{TF}}\left(\mathbf{r}\right)=0$ for $\mathbf{r}\notin\mathbb{D}_{\mathrm{TF}}$.
Here $n_{0}$ is the central density and $\lambda_{x}$, $\lambda_{y}$
and $\lambda_{z}$ are the semi-axes of the ellipsoid. In the case
where only contact interaction is present it is easy to see that $n_{\mathrm{TF}}\left(\mathbf{r}\right)$
is of the form (\ref{eq:nTF}). However, this is not so obvious for
the case of a dipolar BEC. As has been discussed by Eberlein et al.
\cite{Eberlein} the BEC density distribution remains of ellipsoidal
shape also in the presence of the dipole-dipole interaction only the
semi axes being modified. They have also shown that the BEC may become
unstable if the dipole-dipole interaction becomes too large. The dimensionless
parameter $\varepsilon_{D}=\frac{g_{D}}{3g_{s}}$ provides a measure
for the strength of the dipole-dipole interaction compared to the
strength of the contact interaction. In the region $-1/2<\varepsilon_{D}<1$
the ground state is stable while beyond this region it may or may
not be stable depending on the trap geometries. In the following we
will only consider values of $\varepsilon_{D}$ in the stable region.
In the case where no dipole-dipole interaction is present ($\varepsilon_{D}=0$)
the semi axes are given by 
\begin{equation}
\lambda_{a}^{(0)}=\sqrt{\frac{2\mu^{(0)}}{M\omega_{a}^{2}}}\quad a\in\left\{ x,y,z\right\} ,\label{eq:non-dipolar semi axis}
\end{equation}
with the chemical potential $\mu^{(0)}=g_{s}n_{0}^{(0)}$ and $n_{0}^{(0)}$
being the central density of a non-dipolar BEC fixed by the normalization
condition $\int\mathrm{d}\mathbf{r}\, n_{\mathrm{TF}}=N$ and given
by \cite{Pitaevskii&Stingari,Pethick&Smith} 
\begin{equation}
n_{0}^{(0)}=\frac{15}{8\pi}\frac{N}{\lambda_{x}^{(0)}\lambda_{y}^{(0)}\lambda_{z}^{(0)}}.\label{eq:central density}
\end{equation}
For a dipolar BEC these quantities need to be determined numerically,
see for example \cite{Eberlein,Lahaye,Sapina}. Using $n\left(\mathbf{r}\right)=n_{\mathrm{TF}}\left(\mathbf{r}\right)$
integral \eref{eq:Usc} cannot be solved analytically. However,
if the distance $x_{d}$ between the BEC and the superconductor is
large enough, i.e. $\lambda_{x},\lambda_{y}\ll x_{d}$, the problem
can be further simplified. Density distribution \eref{eq:nTF} can
be integrated over $x$ and $y$ yielding 
\begin{equation}
n_{1\mathrm{D}}\left(z\right)=\frac{15}{16}\frac{N}{\lambda_{z}}\left(1-\frac{z^{2}}{\lambda_{z}^{2}}\right)^{2}\label{eq:n1d}
\end{equation}
for $\left|z\right|\leq\lambda_{z}$ and $n_{1\mathrm{D}}\left(z\right)=0$
elsewhere. $n_{1\mathrm{D}}\left(z\right)$ is the so-called column
density \cite{Antezza} and represents an effective one dimensional
density distribution of the mirror BEC. Note that $n_{1\mathrm{D}}\left(z\right)$
is a good approximation for the 3D mirror BEC even for $\lambda_{z}\gg x_{d}$.
Using $n_{1\mathrm{D}}\left(z\right)$ the interaction potential \eref{eq:Usc}
is reduced to an one dimensional integral along the axis of the column
density of the mirror BEC. The potential generated by the mirror BEC
at a position $\mathbf{r}=(x,y=0,z)^{T}$ can be written as 
\begin{equation}
\fl U_{SC}\left(x,z\right)=-\,\frac{g_{D}}{4\pi}\intop_{-\lambda_{z}}^{\lambda_{z}}\mathrm{d}z^{\prime}\, n_{1D}\left(z^{\prime}\right)\left(\frac{3\left(z^{\prime}+z\right)^{2}}{\left(x^{2}+\left(z^{\prime}+z\right)^{2}\right)^{5/2}}-\frac{1}{\left(x^{2}+\left(z^{\prime}+z\right)^{2}\right)^{3/2}}\right).\label{eq:Usc approximation}
\end{equation}
Here, we have evaluated $U_{SC}$ only in the plane $y=0$, since
the column density of the BEC is located in this plane and we are
interested in oscillations in $x$-direction (see next section). The
analytical solution of this integral is straight forward. The discussed
model is depicted in Fig. \ref{Fig01}. 

\section{Center-of-mass frequency shift}

In this section we will discuss the change of the center-of-mass frequency
of the dipolar BEC due to the presence of the superconductor. The
external trapping potential $V_{T}\left(\mathbf{r}\right)$ provides
a certain oscillation frequency which is modified by the interaction
with the mirror BEC. In particular, we are interested in the frequency
of the center-of-mass motion perpendicular to the surface. The frequency
shift due to the superconductor is related to the curvature in $x$-direction
generated by $U_{\mathrm{SC}}$. By calculating the curvature we have
to take into account that the motion of the BEC also leads to motion
of the mirror BEC. When the BEC moves towards the superconductor,
so does the mirror BEC. This means that we have to take the derivative
with respect to the distance to the superconductor rather than with
respect to the distance to the mirror BEC. Using expression \eref{eq:Usc
approximation} as the interaction potential we have to take the derivative
with respect to $x/2$ and the curvature change along the $z$-axis
of the BEC reads 
\[
g\left(z;\, x_{d}\right)=4\left.\frac{\partial^{2}}{\partial x^{2}}U_{SC}\left(x,z\right)\right|_{x=2x_{d}}.
\]
For small amplitude oscillations the center-of-mass frequency $\omega_{x}^{\prime}$
perpendicular to the surface is determined by \cite{Antezza} 
\begin{equation}
\omega_{x}^{\prime2}=\omega_{x}^{2}+\frac{1}{M}\frac{1}{N}\intop_{-\lambda_{z}}^{\lambda_{z}}\mathrm{d}z\, n_{1D}\left(z\right)g\left(z;\, x_{d}\right),\label{eq:frequency change}
\end{equation}
here $n_{1D}\left(z\right)$ represents the column density of the
BEC and $\omega_{x}$ is the frequency of the harmonic trapping potential
$V_{T}\left(\mathbf{r}\right)$. If the frequency change is small
we have $\omega_{x}^{\prime2}-\omega_{x}^{2}=\left(\omega_{x}^{\prime}-\omega_{x}\right)\left(\omega_{x}^{\prime}+\omega_{x}\right)\approx\left(\omega_{x}^{\prime}-\omega_{x}\right)2\omega_{x}$
and with that the relative frequency shift can be written as 
\begin{equation}
\gamma=\frac{\omega_{x}^{\prime}-\omega_{x}}{\omega_{x}}=\frac{1}{2M\omega_{x}^{2}}\frac{1}{N}\intop_{-\lambda_{z}}^{\lambda_{z}}\mathrm{d}z\, n_{1D}\left(z\right)g\left(z;\, x_{d}\right).\label{eq:gamma}
\end{equation}
Using the semi-axes of a non-dipolar BEC $\lambda_{a}=\lambda_{a}^{(0)}$
we find $\gamma\propto\varepsilon_{D}$. The change of the semi-axes
of the ellipsoidal BEC due to the dipole-dipole interaction only appears
as a higher order correction. The integral in expression \eref{eq:gamma}
is best evaluated numerically. Although the atoms in the BEC do not
experience an individual frequency shift, we can still consider a
single atom at the center of the BEC interacting with the mirror BEC
in order to get an analytical order of magnitude estimate for the
frequency shift. In this case we have $\gamma=\frac{g\left(z=0;\, x_{d}\right)}{2\omega_{x}^{2}M}$.
The sign of the dipole-dipole interaction depends on the relative
position of the interacting dipoles, it can be attractive or repulsive.
Considering a single dipole at $z=0$, the strongest frequency shift
is obtained when the interaction sign is the same with all the dipoles
in the mirror BEC (see Fig. \ref{Fig01}). Then, all contributions
to the interaction integral \eref{eq:Usc approximation} add up
constructively. Depending on the distance $x_{d}$ there is an optimal
length of the BEC, which reads for the semi-axis $\lambda_{z}=\sqrt{2}x_{d}$.
Using $\lambda_{z}^{(0)}=\sqrt{2}x_{d}$ we find analytically 
\begin{equation}
\gamma_{\mathrm{max}}=\frac{g\left(z=0;\, x_{d}\right)}{2\omega_{x}^{2}M}\approx0.11\cdot\varepsilon_{D}\left(\frac{\lambda_{x}^{(0)}}{x_{d}}\right)^{4},\label{eq:rule of thumb}
\end{equation}
which represents a rule of thumb for the magnitude of the maximal
frequency shift. As an example, for $^{52}$Cr with $\varepsilon_{D}\approx0.15$
\cite{Chrom2} assuming a distance of $x_{d}=2\lambda_{x}^{(0)}$
we find $\gamma_{\mathrm{max}}\approx10^{-3}$. A frequency shift
of this magnitude is well within experimental resolution. A precision
of $10^{-5}$ was demonstrated in an experiment where the Casimir-Polder
force was measured via the frequency shift of a BEC \cite{Harber}.

\begin{figure}[t]
\flushright\includegraphics{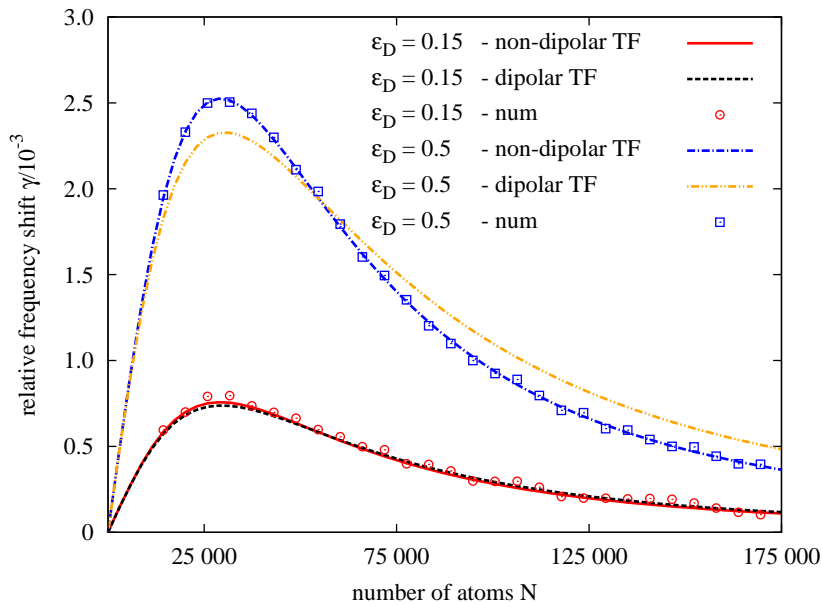} 
\caption{\label{Fig02}The frequency shift as function of the number of atoms
for different models and interaction strengths $\varepsilon_{D}=0.15$
and $\varepsilon_{D}=0.5$. The curves labeled with ``non-dipolar
TF'' represent calculations based on a TF density distribution for
a BEC without dipole-dipole interaction. The curves labeled with ``dipolar
TF'' are based on a TF density distribution which takes the dipole-dipole
interaction into account. In both cases the frequency shift was calculated
using equation \eref{eq:gamma}. The points labeled with ``num''
are the results of a numerical time-evolution of the GPE using the
effective potential \eref{eq:effective potential}. The parameters
for the curves are $\lambda_{x}^{(0)}=\lambda_{y}^{(0)}=\unit[7]{\mu m}$,
$x_{d}=\unit[14]{\mu m}$, $n_{0}^{(0)}=\unit[2.5\times10^{13}]{cm^{-3}}$.
And the semi axis in $z$-direction varies from $\lambda_{z}^{(0)}\approx\unit[7]{\mu m}$
at $N=15\,000$ and $\nu\approx1$ to $\lambda_{z}^{(0)}\approx\unit[85]{\mu m}$
at $N=175\,000$ and $\nu\approx12$.}
\end{figure}

In Fig. \ref{Fig02} we show the frequency shift \eref{eq:gamma}
as a function of the number of atoms $N$ in the BEC. For the calculations
we assumed a distance of $\unit[x_{d}=14]{\mu m}$. Experimental findings
\cite{Kasch,Bernon} and theoretical analysis \cite{Markowsky,Dikovsky} 
both suggest
that using a superconducting microtrap such distances can be achieved.
The frequency shift is presented for two different values of the dipole-dipole
interaction parameter using three different models. First we calculated
the frequency shift \eref{eq:gamma} for a non-dipolar BEC, meaning
that we used the semi-axes $\lambda_{a}^{(0)}$ according to \eref{eq:non-dipolar
semi axis}. So the dipole-dipole interaction is only taken into account
in the interaction between the BEC and its mirror. In Fig. \ref{Fig02}
this model is labeled by ``non-dipolar'' in the plot legend. In
this case the ratio of the trapping frequencies $\nu=\omega_{x}/\omega_{z}$
is given by the inverse ratio of the semi axes $\nu=\lambda_{z}^{(0)}/\lambda_{x}^{(0)}$
according to \eref{eq:non-dipolar semi axis}. The larger the value
of $\nu$ the more elongated is the BEC in $z$-direction. If the
semi axes $\lambda_{x}^{(0)}$ and $\lambda_{y}^{(0)}$ and the central
density $n_{0}^{(0)}$ are kept constant the trap ratio $\nu$ becomes
proportional to the number of atoms, which is easy to see using relation
\eref{eq:central density} 
\begin{equation}
\nu=\frac{\lambda_{z}^{(0)}}{\lambda_{x}^{(0)}}=\frac{15}{8\pi}\frac{N}{n_{0}^{(0)}\left(\lambda_{x}^{(0)}\right)^{2}\lambda_{y}^{(0)}}.\label{eq:nu}
\end{equation}
Varying the number of atoms this way is equivalent to changing length
of the BEC. Experimentally this could be achieved by adjusting the
trap frequency $\omega_{z}$ according to the number of atoms $N$,
such that relation \eref{eq:nu} remains satisfied. Fig. \ref{Fig02}
shows that $\gamma$ has a maximum. This maximum appears at an optimal
length of the BEC as has been discussed above. Increasing the number
of atoms above the optimal number leads to a smaller frequency shift,
because contributions from the edges of the mirror BEC with opposite
sign compensate contributions from the central region. In the limit
$N\rightarrow\infty$ the frequency shift approaches $0$. In order
to detect the eddy current effect experimentally, we suggest to use
the frequency $\bar{\omega}_{x}$ of a long BEC as reference frequency.
If the BEC is long enough, the frequency shift due to the eddy current
effect is smaller than experimentally detectable. However, $\bar{\omega}_{x}$
would still include possible shifts due to other effects, like for
example the Casimir-Polder force, which do not depend on $N$. With
$\bar{\omega}_{x}$ as a reference the frequency shift $\frac{\omega_{x}^{\prime}-\bar{\omega}_{x}}{\bar{\omega}_{x}}$
can be measured as a function of $N$. Since other surface forces
do not have this characteristic dependence on the number of atoms
the curve is a fingerprint for the eddy current effect.

We also calculated the frequency shift using a dipolar BEC. In Fig.
\ref{Fig02} these results are labeled with ``dipolar''. Here we
have used the same parameters as in the other calculation, meaning
that the trap frequencies remain the same as well as the distance
to the surface. However, this time we used the correct dipolar semi
axis $\lambda_{z}$ instead of $\lambda_{z}^{(0)}$ in the density
distribution. While the trap ratio $\nu$ is still proportional to
$N$, it is no longer given exactly be the ratio $\lambda_{z}/\lambda_{x}$.
Also the central density $n_{0}$ slightly changes while varying $N$.
Fig. \ref{Fig02} shows that the effect of the modified semi axes
is negligible for $\varepsilon_{D}=0.15$ but somewhat changes the
result for $\varepsilon_{D}=0.5$. However, the main features of the
curve are preserved.

The points labeled with ``num'' in the plot legend of Fig. \ref{Fig02}
are the results of numerical calculations. We obtained these results
by solving the time dependent GPE \eref{eq:TimeDepGPE} numerically
in three spatial dimensions using a time-splitting spectral method
\cite{TSSP}. We only considered $U_{s}\left(\mathbf{r},\mathbf{r}^{\prime}\right)$
in the GPE for the numerical calculations and will discuss the effect
of $U_{\mathrm{md}}\left(\mathbf{r},\mathbf{r}^{\prime}\right)$ below.
As potential in the GPE we used the following effective potential
\begin{equation}
V_{\mathrm{eff}}\left(\mathbf{r}\right)=\frac{M}{2}\left[\omega_{x}^{2}\left(1+f\left(z;\, x_{d}\right)\right)^{2}x^{2}+\omega_{y}^{2}y^{2}+\omega_{z}^{2}z^{2}\right],\label{eq:effective potential}
\end{equation}
where the function $f(z;\, x_{d})$ describes the relative curvature
change of the potential in $x$-direction due to $U_{SC}$ and is
defined by 
\begin{equation}
f\left(z;\, x_{d}\right)=\frac{1}{2M\omega_{x}^{2}}g\left(z;\, x_{d}\right).\label{eq:f(z,xd)}
\end{equation}
The effective potential \eref{eq:effective potential} is a good
approximation if $f\left(z;\, x_{d}\right)\ll1$. We first determine
the ground state of the GPE numerically. To excite the center-of-mass
oscillation we then shift the potential by a distance $x_{s}$ in
$x$-direction, and calculate the time evolution. In every time step
$t_{i}$ we calculate the $x$-coordinate of the center-of-mass
\[
\left\langle x\left(t_{i}\right)\right\rangle =\intop\mathrm{d}\mathbf{r}\, x\cdot\left|\psi\left(\mathbf{r};\, t_{i}\right)\right|^{2},
\]
where $\psi\left(\mathbf{r};\, t_{i}\right)$ is the numerically determined
solution of the GPE at that particular time step. After the time evolution
is completed we perform a Fourier analysis of the data to obtain the
oscillation frequency. The results for the center-of-mass frequency
are presented in Fig. \ref{Fig02} and labeled ``num''. We can see
a very nice agreement with the results obtained for $\gamma$ using
\eref{eq:gamma}. The reason that it does not agree with the results
for the dipolar BEC is that we did not take into account the modified
semi axes when we calculated $V_{\mathrm{eff}}\left(\mathbf{r}\right)$.
Since we neglected the dipole-dipole interaction in the GPE, neglecting
it in $f(z;\, x_{d})$ is consistent. Again the dipole-dipole interaction
is only taken into account between the BEC and its mirror. The numerically
obtained results are expected to follow the curves labeled ``dipolar''
in Fig. \ref{Fig02} if we include the dipole-dipole interaction in
the GPE and consider it in $f(z;\, x_{d})$.

\section{Coupling of the center-of-mass motion with the breather mode}

Next we want to discuss the aforementioned coupling of a collective
mode with the center-of-mass motion due to the eddy current effect.
Center-of-mass motions can be excited by a sudden shift $x_{s}$ of
the trap minimum. In a harmonic potential the center-of-mass motion
does not excite collective shape fluctuations of the BEC and the shape
of the BEC will remain constant during motion. However, in the vicinity
of the superconductor the effective potential is no longer purely
harmonic. The interaction with the mirror BEC generates additional
anharmonic terms to the harmonic trapping potential. The excitation
of collective modes due to terms like $x^{3}$, $x^{4}$ etc. has
been discussed previously \cite{Anha1,Ott,Anha2,Anha3}. The lowest
order anharmonic term in \eref{eq:effective potential} is of the
form $x^{2}z^{2}$. Transforming into the center-of-mass system we
have $x\left(t\right)\propto\sin\left(\omega_{x}^{\prime}t\right)$.
The anharmonic term thus generates a time dependent change of the
trap frequency in $z$-direction of the form $\Delta\omega_{z}^{\prime}\left(t\right)\propto\sin^{2}\left(\omega_{x}^{\prime}t\right)$.
This excites monopole-quadrupole modes of the BEC \cite{Pitaevskii}
with frequency $2\omega_{x}^{\prime}$. We thus expect to see a resonance,
if one of the monopole-quadrupole modes happens to have the frequency
$2\omega_{x}^{\prime}$. The best candidate for this is the so-called
breather mode.

\begin{figure}
\flushright\includegraphics{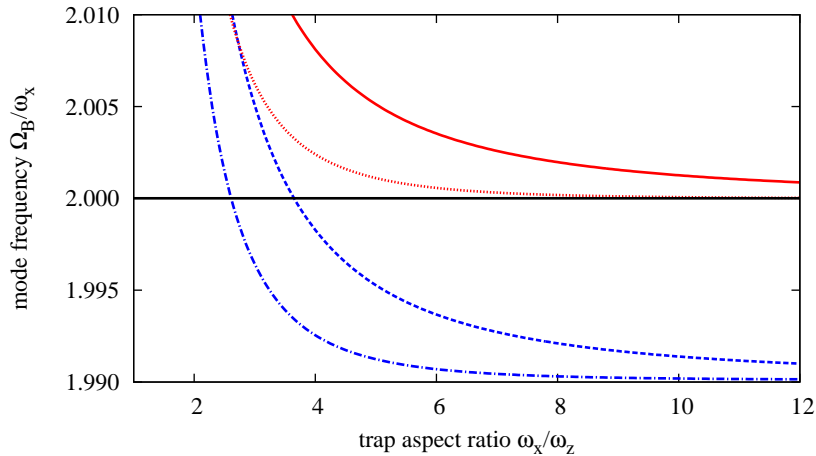} 
\caption{\label{Fig03}Breather mode frequency $\Omega_{B}$ versus $\omega_{x}/\omega_{z}$
for different $\varepsilon_{D}$ and $\omega_{y}/\omega_{x}$: red
solid line $\varepsilon_{D}=0$ and $\omega_{y}/\omega_{x}=1$; red
dotted line $\varepsilon_{D}=0.15$ and $\omega_{y}/\omega_{x}=1$;
blue dashed line $\varepsilon_{D}=0$ and $\omega_{y}/\omega_{x}=0.99$;
blue dot-dashed line $\varepsilon_{D}=0.15$ and $\omega_{y}/\omega_{x}=0.99$;}
\end{figure}

In Fig. \ref{Fig03} we show the breather mode frequency $\Omega_{B}$
of a BEC calculated within the TF approximation \cite{Sapina,Bijnen}.
The mode frequency is shown as a function of the aspect ratio $\omega_{x}/\omega_{z}$
of the trapping frequencies. In a spherical trap with $\omega_{x}=\omega_{y}=\omega_{z}$
the mode frequency is $\Omega_{B}=\sqrt{2}\omega_{x}$. In a uni-axial
elongated trap with $\omega_{x}=\omega_{y}>\omega_{z}$ the mode frequency
approaches $2\omega_{x}$ (red line). In a tri-axial elongated trap
with $\omega_{x}>\omega_{y}>\omega_{z}$ the mode frequency crosses
$2\omega_{x}$ (dashed blue line). In the latter case we expect to
see a strong resonant coupling of the breather mode and the center-of-mass
motion at the crossing point.

To study the excitation of the breather mode, we again numerically
solve the time dependent GPE using potential \eref{eq:effective
potential}. However, this time we analyze the fluctuations of the
width of the wave function 
\begin{equation}
\sigma_{x}=\sqrt{\left\langle \left(x-\left\langle x\right\rangle \right)^{2}\right\rangle }.\label{eq:sigma_x}
\end{equation}
After calculating the time evolution a Fourier analysis of $\sigma_{x}$
provides the frequency spectrum. The peaks in the frequency spectrum
correspond to collective modes of the BEC. We have identified the
peaks in the spectrum by comparing them to the results from TF calculations.

\begin{figure}
\flushright\includegraphics{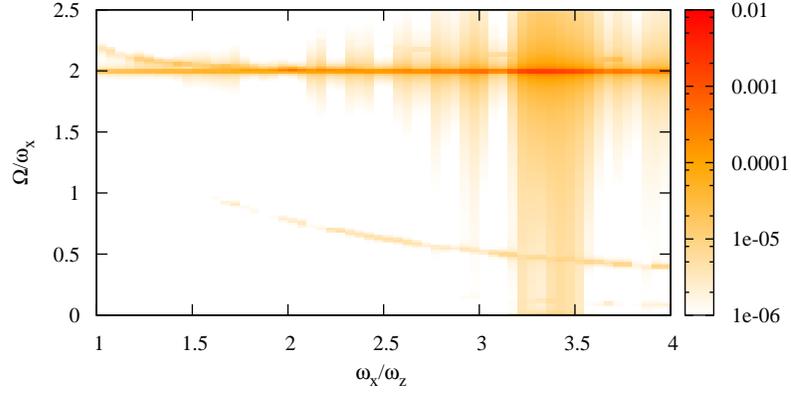}
\caption{\label{Fig04}Numerical results for the excitation spectrum of $\Delta\sigma_{x}$
for different trap ratios $\omega_{x}/\omega_{z}$. The trap ratios
range from $\omega_{x}/\omega_{z}=1$ to $\omega_{x}/\omega_{z}=4$
in steps of $0.05$. The colors represent the amplitudes on a logarithmic
scale. The mode frequencies range from $\Omega/\omega_{x}=0$ up to
$\Omega/\omega_{x}=2.5$. There are two modes visible, which correspond
to collective modes of the BEC. The third structure at $\approx2\omega_{x}$
corresponds to twice the oscillation frequency of the center-of-mass
which excites the collective modes. Parameters: $\omega_{y}/\omega_{x}=0.99$;
$\varepsilon_{D}=0.15$; length of time evolution $t=500\, T_{x}$,
with $T_{x}=\frac{2\pi}{\omega_{x}}$.}
\end{figure}

In Fig. \ref{Fig04} we present the frequency spectra of $\Delta\sigma_{x}=\sigma_{x}\left(t\right)-\sigma_{x}\left(0\right)$
in a slightly tri-axial trap for different $\omega_{x}/\omega_{z}$.
The yellow line starting at $\approx2.3\,\omega_{x}$ represents the
breather mode frequency. For increasing trap ratio $\omega_{x}/\omega_{z}$
it approaches another yellow line at $\approx2\omega_{x}$. This second
line represents the double oscillation frequency of the center-of-mass
motion. A resonance can be observed where the two lines meet (red). There
is also a third yellow line visible in the spectrum. It belongs to
another collective mode of the BEC, which also is excited due to the
anharmonicity of the trap. However, this excitation is very weak compared
to the resonance which occurs for the breather mode.

\begin{figure}
\flushright\includegraphics{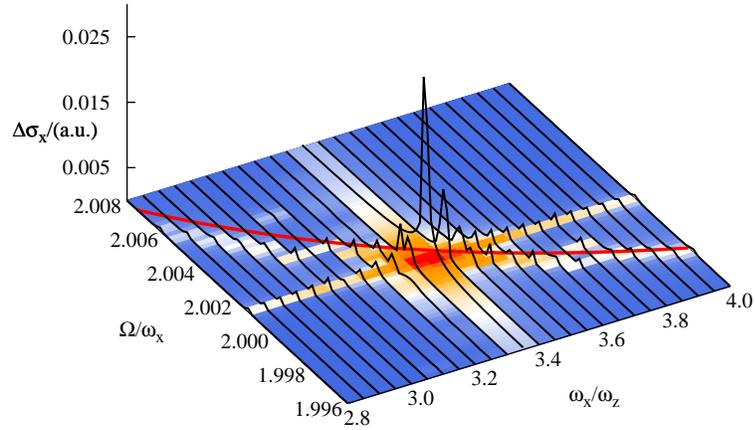}
\caption{\label{Fig05}Numerical results for the excitation spectrum of $\Delta\sigma_{x}$
for different trap ratios $\omega_{x}/\omega_{z}$. In this figure
an enlarged region near the resonance in Fig. \ref{Fig04} is shown.
Parameters: $\omega_{y}/\omega_{x}=0.99$; $\varepsilon_{D}=0.15$;
length of time evolution $t=5\times10^{3}\, T_{x}$, with $T_{x}=\frac{2\pi}{\omega_{x}}$.
The red line represents $\Omega_{B}$ within TF approximation for
a non-dipolar BEC (blue dashed line in Fig. \ref{Fig03}).}
\end{figure}

In Fig. \ref{Fig05} an enlarged region of the excitation spectrum
of $\Delta\sigma_{x}$ near the resonance is shown. While the spectra
are plotted on a linear scale the color map on the bottom is plotted
on a logarithmic scale. There are two peaks visible in the spectra
shown: one corresponding to the breather mode frequency and the other
representing twice the center-of-mass frequency ($\approx2.0012\,\omega_{x}$).
One can clearly see the resonance peak at the crossing point just
as we have already conjectured.

As shown in Fig. \ref{Fig03} the breather mode for a BEC with $\varepsilon_{D}>0$
has a different frequency than a BEC with $\varepsilon_{D}=0$. Since
we have neglected $U_{\mathrm{md}}\left(\mathbf{r},\mathbf{r}^{\prime}\right)$
in the GPE this effect is not included in our numerical results. Fig.
\ref{Fig03} shows that the crossing point of the breather mode and
the double oscillation frequency is shifted towards smaller values
of $\omega_{x}/\omega_{z}$ (dot-dashed blue line). Therefore we also
expect the resonance peak to appear at a smaller value of $\omega_{x}/\omega_{z}$.

\section{Discussion and Conclusion}

We have studied a new physical mechanism to couple a BEC and a superconductor.
The mechanism rests on the interaction of the magnetic dipole moments
of a dipolar BEC with the superconductor. Center-of-mass oscillations
of such a dipolar BEC create eddy currents in the superconductor,
which act back on the BEC. We have demonstrated that this eddy current
effect leads to a frequency change and can resonantly excite a collective
mode of the BEC. 

The frequency change of the center-of-mass motion of the BEC has a
characteristic dependence on the number of atoms in the BEC, which
can serve as an experimental fingerprint to distinguish this effect
from other effects that may change the oscillation frequency. We have
also shown that the resonant excitation of the collective mode of
the BEC becomes possible, if the parameters of the external trapping
potential are tuned properly. Both effects become significant in $^{52}$Cr,
$^{168}$Er or $^{164}$Dy BECs.

In principle there should be similar effects using a thermal cloud
instead of a BEC. However, it will be much harder to observe those
effects, because the density in a thermal cloud is much smaller. Also
such a system would not represent a coupled quantum system. 

The eddy current effect we described here requires a trap to
be formed a few ten micrometers away from a superconductor
surface. For superconducting microtraps it has been pointed out
that the Meissner effect of the superconductor modifies
the magnetic field distribution in its vicinity in such
a way that a magnetic trap cannot be formed anymore,
if the trap position is brought too close to the 
superconducting surface \cite{CanoPRL}.
In Refs. \cite{Markowsky} and \cite{Dikovsky} 
this effect has been studied theoretically
for a finite conductor with rectangular cross-section and
the limits for such traps have been discussed. 
It was shown that a trap distance of less than $\unit[10]{\mu m}$
can be achieved for a Niobium conductor. A recent experiment 
\cite{Bernon} has demonstrated a distance of 
$\unit[14]{\mu m}$ from a superconducting strip.

In our calculations we assumed a superconducting half-space.
For this approximation to be appropriate a finite superconducting strip 
with a rectangular cross-section should fulfil the
following requirements:
\begin{itemize}
\item Thickness: as the eddy currents are flowing at the surface of the superconductor
within the magnetic penetration depth $\lambda$, the thickness of the 
superconductor should be $2\lambda$ or more, which for Niobium is $\sim$200~nm. 
These are typical film thicknesses used in thin film technology. 
\item Width: the width of the strip should be larger than both the 
width of the BEC and its distance from the superconductor.
In the recent experiment of Ref. \cite{Bernon} it was 
shown that this requirement can be met.
\item Length: the length should be larger than both the length of the BEC 
and its distance from the superconductor. Such an axial confinement 
could be made by a Z-shape trap, for example, as in the work of
Ref. \cite{Bernon}.
\end{itemize}

If some of these requirements cannot be met in an experiment,
it is clear that the eddy current effect we describe will be
reduced by a geometrical factor. This factor depends on the
solid angle under which the BEC sees the superconductor surface.

In this work we have only considered small amplitude oscillations.
For large amplitudes we expect to see corrections to the results presented
here. On the other hand, larger amplitudes should increase the discussed
effects since the potential becomes more anharmonic. For example the
potential can no longer be assumed to be symmetric in the $x$-direction.

So far experiments on superconducting microtraps for BECs have only
been done with $^{87}$Rb atoms, which have a comparatively small
magnetic dipole moment. Our calculations show that it is beneficial
to study strongly dipolar BECs in superconducting microtraps. In such
a system a mutual coupling of a BEC and a superconductor via eddy
currents becomes possible.

\section*{Acknowledgments}

We acknowledge useful discussions with J. Fort\'agh, H. Hattermann,
and N. Schopohl. IS would like to thank the Physics Departments of
the University of Hull and the University of Bielefeld for their hospitality
during stays at which part of this work has been done. This work was
supported by the Deutsche Forschungsgemeinschaft via SFB/TR21.

\end{document}